\documentclass[american,aps,prl,reprint,superscriptaddress,longbibliography]{revtex4-2}
\usepackage{lmodern}
\usepackage[T1]{fontenc}
\usepackage[latin9]{inputenc}
\usepackage{color}
\usepackage{babel}
\usepackage{amsmath}
\usepackage{amssymb}
\usepackage{graphicx}
\usepackage{esint}
\usepackage[unicode=true,pdfusetitle,
 bookmarks=true,bookmarksnumbered=false,bookmarksopen=false,
 breaklinks=false,pdfborder={0 0 0},pdfborderstyle={},backref=false,colorlinks=true]
 {hyperref}
\hypersetup{
 citecolor=blue,linkcolor=magenta,urlcolor=blue}

\makeatletter
\usepackage{xcolor}

\makeatother

\begin{document}
\global\long\def\i{\mathrm{i}}%
\global\long\def\e{\mathrm{e}}%
\global\long\def\d{\mathrm{d}}%
\global\long\def\bra#1{\left\langle #1\right|}%
\global\long\def\ket#1{\left|#1\right\rangle }%
\global\long\def\braket#1#2{\left\langle #1|#2\right\rangle }%
\global\long\def\ketbra#1#2{\left|#1\right\rangle \!\left\langle #2\right|}%
\global\long\def\Tr{\mathrm{Tr}}%

\title{\textcolor{black}{Eight-dimensional topological systems simulated
using time-space crystalline structures}}
\author{Yakov Braver}
\affiliation{Institute of Theoretical Physics and Astronomy, Vilnius University,
Saul\.{e}tekio 3, LT-10257 Vilnius, Lithuania}
\author{Egidijus Anisimovas}
\affiliation{Institute of Theoretical Physics and Astronomy, Vilnius University,
Saul\.{e}tekio 3, LT-10257 Vilnius, Lithuania}
\author{Krzysztof Sacha}
\affiliation{Instytut Fizyki Teoretycznej, Uniwersytet Jagiello\'{n}ski, ulica
Profesora Stanis\l awa \L ojasiewicza 11, PL-30-348 Krak\'{o}w, Poland}
\begin{abstract}
We demonstrate the possibility of using time-space crystalline structures
to simulate eight-dimensional systems based on only two physical dimensions.
A suitable choice of system parameters allows us to obtain a gapped
energy spectrum, making topological effects become relevant. The nontrivial
topology of the system is evinced by considering the adiabatic state
pumping along temporal and spatial crystalline directions. Analysis
of the system is facilitated by rewriting the system Hamiltonian in
a tight-binding form, thereby putting space, time, and the additional
synthetic dimensions on an equal footing.
\end{abstract}
\maketitle
\emph{Introduction}. Quantum simulation is a rapidly growing and exciting
field of study, focused on exploiting controllable quantum systems
to replicate and probe complex physical phenomena \citep{Feynman1982,fraxanet2022coming}.
Whereas simulating a quantum system with full precision is a daunting
task, recreation of specific relevant features is often within reach.
In particular, ultracold atomic systems in optical lattices \citep{schafer2020tools}
have successfully demonstrated abilities to model intricate condensed-matter
and topological phenomena \citep{Chiu2019,ozawa2019topological} as
well as lattice gauge theories \citep{Banuls2020simulating,Aidelsburger2021coldmeet}.
An intriguing line of thought along this direction is the emulation
of high-dimensional systems, notably --- high-dimensional periodically
ordered physical structures --- in low-dimensional settings \citep{PetridesEtAl2018,LeeEtAl2018,ozawa2019topological,Price2020,Zlabys2021,Zhu2022topo4d}.
In this context, several recent proposals drew inspiration from the
emergent concept of time crystals \citep{Wilczek2012,Shapere2012,GuoBook,SachaTC2020}
and asked if time can play the role of an additional coordinate in
quantum simulations. This time-crystalline approach \citep{Sacha2017rev,SachaTC2020}
involves a driving signal of a certain frequency to create a repeating
pattern of motion at a commensurate frequency that persists over time.
Many condensed matter phenomena were thus reenacted in the time domain
\citep{SachaTC2020,Hannaford2022}, and the possibility to engage
both temporal and spatial dimensions at the same time was established
\citep{Li2012,Gao2021,Zlabys2021,Braver2022}, thus doubling the number
of available dimensions.

In this Letter, we provide a route for studying topological eight-dimensional
(8D) systems that can be experimentally realized using only two physical
spatial dimensions. We start with a periodically driven 1D optical
lattice with steep barriers (modeled by delta-functions) and show
that it can sustain a 2D time-space lattice. The topological nature
of the attained time-space crystalline structure is made evident by
considering adiabatic state pumping along temporal and spatial crystalline
directions. Interpreting the two adiabatic phases as crystal momenta
of simulated extra dimensions, we show that the energy bands of the
system are characterized by nonvanishing second Chern numbers of the
effective 4D lattice. Finally, we demonstrate that two such 4D systems
can be combined, and the resulting energy spectrum will remain gapped.
The topological properties of the attained 8D system are then characterized
by the fourth Chern number, and energy bands with nonvanishing values
of the fourth Chern number are identified.

\begin{figure}
\begin{centering}
\includegraphics{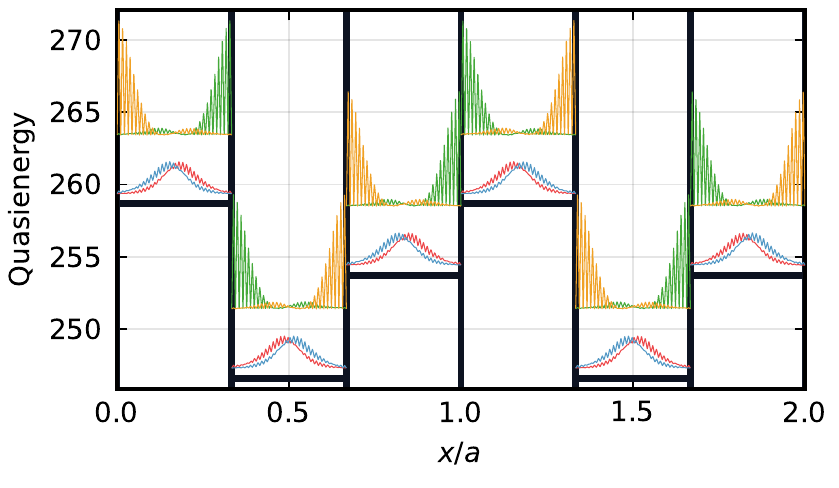}
\par\end{centering}
\caption{\label{fig:potential}Wannier function densities $|w_{\ell}(x,t=7\pi/4\Omega)|^{2}$
at $\varphi_{x}=\pi/5$, $\varphi_{t}=0$, where $2\pi/\Omega$ is
the period of time evolution of $w_{\ell}(x,t)$. The baseline of
each Wannier function is shifted upwards along the quasienergy axis
by the mean value of quasienergy in the corresponding state, i.e.
the quantity $\protect\bra{w_{\ell}}\hat{{\cal H}}\protect\ket{w_{\ell}}$.
At the chosen value of $\varphi_{x}$, each Wannier function is localized
almost entirely within a single site, therefore, the infinitesimal
\textquotedblleft tails\textquotedblright{} of the functions are not
shown. Black vertical lines represent the steep barriers separating
the sites of the spatial lattice, while the horizontal lines depict
the values of the onsite energies described by the third term in Eq.~(\ref{eq:h}).
Parameters of the model are: $N=2$, $s=2$, $a=4.000$, $V=2000$,
$U=7.000$, $\omega=676.8$, $\lambda_{{\rm S}}=10.00$, $\lambda_{{\rm L}}=20.00$.
Trailing zeros are within the numerical resolution/accuracy and are
significant. The nature of the presented results, however, will not
change if all the values are chosen within $\sim10\%$ of the given
ones and then $\omega$ is adjusted accordingly to ensure that the
quasienergy spectrum is gapped.}
\end{figure}

\emph{Model}. We introduce a 1D time-dependent Hamiltonian

\begin{equation}
\hat{H}(x,\hat{p}_{x},t|\varphi_{x},\varphi_{t})=\hat{h}(x,\hat{p}_{x}|\varphi_{x})+\xi(x,t|\varphi_{t}),\label{eq:H}
\end{equation}
written as a sum of an adiabatic-pumping part $\hat{h}$ (which is
static but depends on a spatial adiabatic phase $\varphi_{x}$) and
a time-periodic driving term featuring a second adiabatic phase $\varphi_{t}$.
Throughout this work, we use the recoil units for the energy $\hbar^{2}k_{{\rm L}}^{2}/2m$
and length $1/k_{{\rm L}}$, with $k_{{\rm L}}$ being the wave number
of the primary laser beam used to create the optical lattice and $m$
the particle mass. The unit of time is $\hbar$ divided by the energy
unit. The first term of Eq.~(\ref{eq:H}) is the unperturbed spatial
Hamiltonian,

\begin{equation}
\hat{h}=\hat{p}_{x}^{2}+V\sum_{n=0}^{3N}\delta(x-\tfrac{na}{3})+U\sum_{n=1}^{3}g_{n}(x)\cos\!\left[\varphi_{x}+\tfrac{2\pi(n-1)}{3}\right].\label{eq:h}
\end{equation}
Here, $\hat{p}_{x}$ is the momentum operator, while the sums describe
the spatial potential --- a lattice of $N$ identical cells of length
$a$, each consisting of three sites separated by steep delta-function
barriers, see Fig.~\ref{fig:potential}. The superlattice potential
$g_{n}(x)$ is equal to unity only in the $n$th site of each spatial
cell and vanishes otherwise. This term modulates the onsite energies
in the same way in each cell by changing $\varphi_{x}$, with $U$
controlling the modulation amplitude. Note that the modulation phase
in each consecutive site is lagging with respect to its neighbor on
the left by one third of a cycle. If the modulation is performed adiabatically,
the Thouless pumping can be realized in the system described by $\hat{h}$.
The realization of sharp optical barriers as well as three-site Thouless
pumping have already been studied in the literature \citep{Lacki2016,Tangpanitanon2016,Haug2019}.

The spatial Hamiltonian is perturbed by the terms
\begin{equation}
\begin{split}\xi(x,t|\varphi_{t})=\  & \lambda_{{\rm S}}\cos(\tfrac{12\pi x}{a})\cos(2\omega t)\\
 & +\lambda_{{\rm L}}\cos(\tfrac{6\pi x}{a})\cos(\omega t+\varphi_{t}),
\end{split}
\label{eq:perturb}
\end{equation}
where $\lambda_{{\rm S}}$ and $\lambda_{{\rm L}}$ control the overall
strength of the perturbation. The spatial frequencies $6\pi/a$ and
$12\pi/a$ ensure that all spatial sites are perturbed in the same
way. The driving frequency $\omega$ is chosen so that a resonant
condition is fulfilled in each spatial site. In the classical description,
the resonance means that $\omega$ is very close to an integer multiple
of the frequency $\Omega$ of the periodic motion of a particle in
a spatial site, i.e., $\omega\approx s\Omega$, where $s$ is integer.
In the quantum description, the resonance corresponds to $\omega$
being close to an integer multiple of the gap $\Omega$ between certain
bands of the Hamiltonian (\ref{eq:h}). In the limit $V\rightarrow\infty$
{[}see Eq.~(\ref{eq:h}){]} an independent time-crystalline structure
is formed in each spatial site due to the resonant driving. Specifically,
in the frame evolving along the resonant trajectory, the resonant
dynamics of a particle can be described by $\hat{H}_{{\rm eff}}=\hat{\tilde{p}}_{\tilde{x}}^{2}+\tilde{\lambda}_{{\rm S}}\cos(2s\tilde{x})+\tilde{\lambda}_{{\rm L}}\cos(s\tilde{x}+\varphi_{t})$
where $\tilde{x}\in[0,2\pi)$, see Refs.~\citep{Zlabys2021,Braver2022}.
For example, for $s=2$, there are two temporal cells, each consisting
of two temporal sites. An adiabatic change of the phase $\varphi_{t}$
allows for a realization of the Thouless pumping in the time-crystalline
structures \citep{Braver2022}. If $V<\infty$, then tunneling of
a particle between spatial sites is possible, and the entire system
forms a 2D time-space crystalline structure which, as we will show,
can be described by a 2D tight-binding model.

To study the emergence of a time-space crystalline structure and the
pumping dynamics, we solve the eigenvalue problem $\hat{{\cal H}}u_{n,k}(x,t)=\varepsilon_{n,k}u_{n,k}(x,t)$
for the Floquet Hamiltonian $\hat{{\cal H}}=\hat{H}-\i\partial_{t}$
\citep{Shirley1965,Buchleitner2002,Holthaus2015}. We assume periodic
boundary conditions for the spatial system and introduce the spatial
quasimomentum $k$. We denote the quasienergy of the $n$th eigenstate
by $\varepsilon_{n,k}$, while $u_{n,k}(x,t)$ is the corresponding
Floquet mode that respects temporal periodicity of the perturbation:
$u_{n,k}(x,t)=u_{n,k}(x,t+2\pi/\omega)$. A general solution of the
Schr�dinger equation can be represented as a superposition of states
$\Psi_{n,k}(x,t)=\e^{-\i\varepsilon_{n,k}t}u_{n,k}(x,t).$ In our
simulations we consider a finite number of spatial cells, $N=2$,
and a finite number of temporal cells, $s=2$. The considered values
of quasimomentum are thus $k=0$ and $k=\pi$ (assuming $k\in[0;2\pi)$),
corresponding to the boundary of the Brillouin zone. Consequently,
the obtained widths of the energy bands coincide with the widths being
approached in the limit $N\to\infty$.

The details of the diagonalization procedure are covered in the Supplemental
Material \citep{supplement}. All calculations have been performed
using a number of software packages \citep{JuliaDiffEq,JuliaDiffEq2,JuliaDiffEqKahanLi,JuliaDiffEqMcAte,JuliaOptim,JuliaIntervalArithmetic}
written in the Julia programming language \citep{Julia}. The source
code of our package is available on GitHub \citep{package}.

The resonant subspace of the entire Hilbert space which we are interested
in consists of $3N\times2s$ eigenstates. Diagonalizing the periodic
position operator $\e^{\i\frac{2\pi}{Na}x}$ in this subspace \citep{Aligia1999,Asboth2016short}
we obtain $6Ns$ Wannier functions $w_{\ell}(x,t)$ of the $3N\times2s$
time-space crystalline structure which are represented by localized
wave packets propagating with the period $2\pi/\Omega$ along the
resonant orbits in each spatial site. These Wannier functions are
shown at $t=7\pi/4\Omega$ in Fig.~\ref{fig:potential}, where each
spatial site hosts $2s=4$ states.

\begin{figure*}
\begin{centering}
\includegraphics{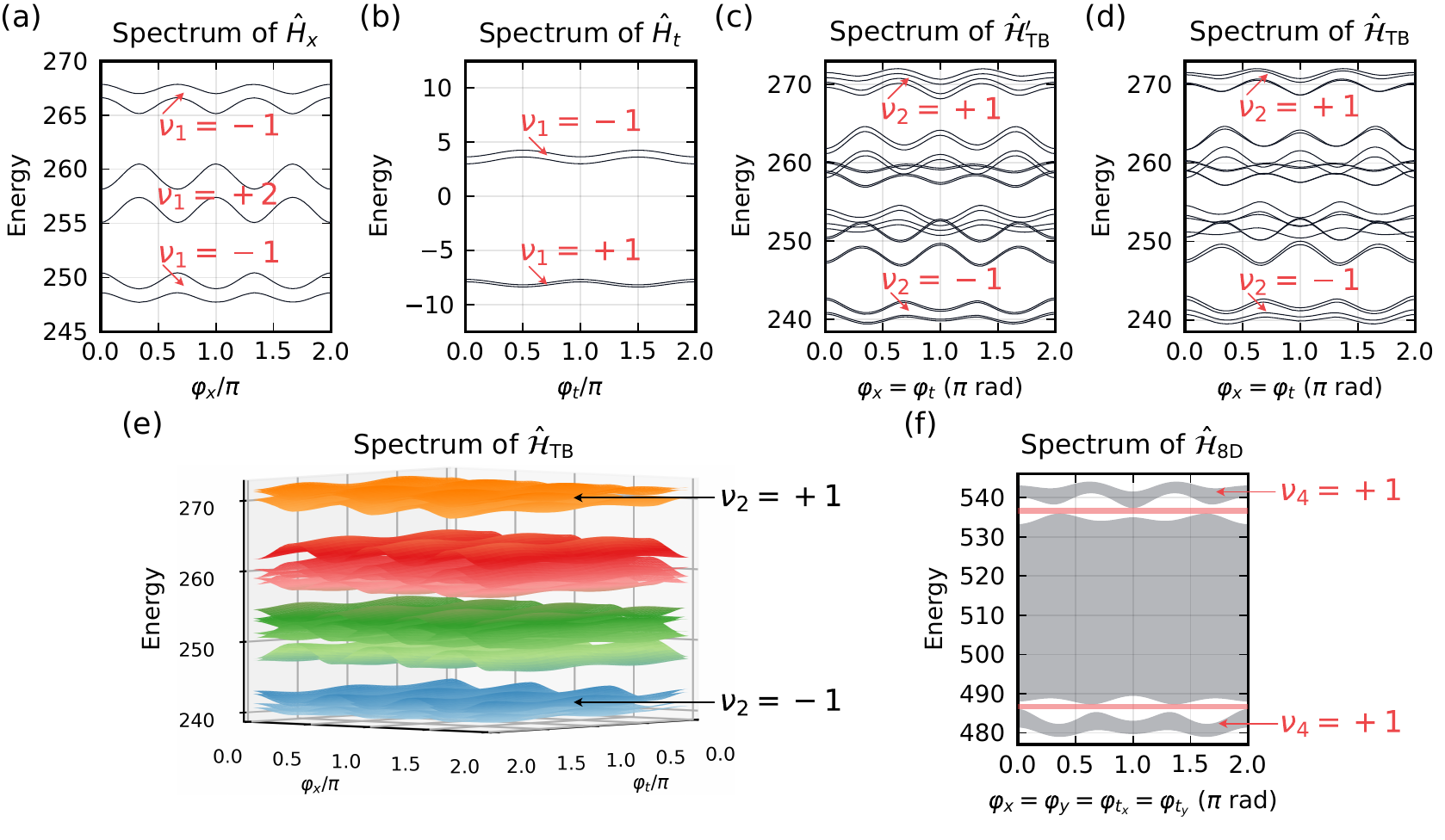}
\par\end{centering}
\caption{\label{fig:spectra}The energy spectra of derived systems. (a) Energy
spectrum of the decomposed spatial Hamiltonian $\hat{H}_{x}$. (b)
Energy spectrum of the decomposed temporal Hamiltonian $\hat{H}_{t}$.
(c) Energy spectrum of $\hat{{\cal H}}'_{{\rm TB}}$, equal to the
Minkowski sum of the spectra in (a) and (b). A cut of the spectrum
along the line $\varphi_{x}=\varphi_{t}$ is shown. (d) Energy spectrum
of $\hat{{\cal H}}{}_{{\rm TB}}$ along the line $\varphi_{x}=\varphi_{t}$.
(e) Eigenenergy surfaces of $\hat{{\cal H}}{}_{{\rm TB}}$. (f) Energy
spectrum of an 8D systems obtained by combining two independent copies
of the 4D systems whose spectra are shown in (e). The gray areas represent
the bands, with individual levels not shown for visual clarity. The
red regions indicate the gaps.}
\end{figure*}

\emph{The tight-binding picture}. In the basis of the Wannier functions,
the Floquet Hamiltonian restricted to the resonant subspace takes
the form of the tight-binding model
\begin{equation}
\hat{{\cal H}}_{{\rm TB}}(\varphi_{x},\varphi_{t})=\sum_{\ell',\ell}J_{\ell'\ell}(\varphi_{x},\varphi_{t})\hat{a}_{\ell'}^{\dagger}\hat{a}_{\ell},
\end{equation}
where operator $\hat{a}_{i}^{\dagger}$ creates (while $\hat{a}_{i}$
annihilates) a boson on site $\ell$. Here, $\ell\in[1,6Ns]$ enumerates
all sites of the 2D time-space lattice, and it is related to the space-time
index pair $(j,\alpha)$ as $\ell=2s(j-1)+\alpha$, where $j\in[1,3N]$
and $\alpha\in[1,2s]$. The matrix elements $J_{\ell'\ell}$ are calculated
as
\begin{equation}
J_{\ell'\ell}=\intop_{0}^{sT}\frac{\d t}{sT}\bra{w_{\ell'}}\hat{{\cal H}}\ket{w_{\ell}},
\end{equation}
where $T=2\pi/\omega$ is the driving period. The Wannier basis is
constructed repeatedly for every phase $\varphi_{x}$ and $\varphi_{t}$.
Each state $\ket{w_{\ell}(t)}$ is confined to a single spatial site,
consequently, only nearest-neighbor spatial couplings are relevant.
Moreover, this coupling is appreciable only at times when a given
state $\ket{w_{\ell}(t)}$ is localized near a classical turning point
(see the green and yellow states in Fig.~\ref{fig:potential}). At
these times, each of these states has only one partner which it is
coupled to. Therefore, each Wannier state is coupled to only a single
state of those in the neighboring spatial sites. Provided these partners
(see like-colored states in Fig.~\ref{fig:potential}) are numbered
with the same temporal index $\alpha$, it will not change when a
state transitions to a neighboring site (only $j$ will change). This
leads to a separable structure of the resulting time-space lattice,
where ``diagonal'' transitions --- those which require both indices
$j$ and $\alpha$ to change simultaneously --- are forbidden. This
is an idealized picture, but one which holds with high accuracy since
next-nearest-neighbor couplings are negligible (see \citep{supplement}).
Note that this separability is intrinsic to the model described by
Eqs.~(\ref{eq:H})--(\ref{eq:perturb}) and cannot be changed by
tuning the parameters.

Thus, the Hamiltonian $\hat{{\cal H}}_{{\rm TB}}$ is separable in
the sense that 
\begin{equation}
\hat{{\cal H}}_{{\rm TB}}\approx\hat{I}_{x}\otimes\hat{H}_{t}+\hat{H}_{x}\otimes\hat{I}_{t}\equiv\hat{{\cal H}}'_{{\rm TB}},\label{eq:decomposition}
\end{equation}
where ``$\otimes$'' denotes the tensor product, $\hat{H}_{x}$
and $\hat{H}_{t}$ are, respectively, the separated spatial and temporal
Hamiltonians, while $\hat{I}_{x}$ and $\hat{I}_{t}$ are the identity
operators acting in the spaces of, respectively, operators $\hat{H}_{x}$
and $\hat{H}_{t}$. Consequently, the eigenvalue spectrum of $\hat{{\cal H}}'_{{\rm TB}}$
is the Minkowski sum of eigenvalue spectra of $\hat{H}_{x}$ and $\hat{H}_{t}$.
We will refer to the eigenvalues of all tight-binding Hamiltonians
as simply ``energies''.

The spectra of $\hat{H}_{x}$ and $\hat{H}_{t}$ are shown in Figs.~\ref{fig:spectra}(a)
and \ref{fig:spectra}(b) together with the first Chern numbers of
each band. Considering the spatial part described by $\hat{H}_{x}$,
we treat the phase $\varphi_{x}$ as a fictitious quasimomentum, allowing
us to introduce the Berry curvature of the $n$th band, $\Omega_{k_{x}\varphi_{x}}=-2\Im\braket{\partial_{k_{x}}\chi_{n,k_{x}}}{\partial_{\varphi_{x}}\chi_{n,k_{x}}}$,
and the corresponding first Chern number \citep{Xiao2010,Nakajima2016,Lohse2016pump,PetridesEtAl2018}
\begin{equation}
\nu_{1}^{(x)}=\frac{1}{2\pi}\int_{{\rm BZ}}\d k_{x}\intop_{0}^{2\pi}\d\varphi_{x}\,\Omega_{k_{x}\varphi_{x}}.\label{eq:nu1}
\end{equation}
In the definition of the Berry curvature, $\ket{\chi_{n,k_{x}}}$
is the cell-periodic part of the Bloch eigenstate $\e^{\i k_{x}j}\chi_{n,k_{x}}(j)$
of $\hat{H}_{x}$ with $\chi_{n,k_{x}}(j+3)=\chi_{n,k_{x}}(j)$ where
$j$ labels spatial sites. For clarity, we suppress indication of
the parametric dependence on $\varphi_{x}$ in $\hat{H}_{x}$ and
its eigenstates. The crystal momentum $k_{x}$ is treated as a continuous
quantity assuming $N\to\infty$. The values of $\nu_{1}^{(x)}$ for
the bands shown in Fig.~\ref{fig:spectra}(a) may be easily determined
as the number of particles of a given band pumped through an arbitrary
lattice cross section per pumping cycle (see \citep{supplement})
or, equivalently, by counting the number of edge state branches in
the spectrum of the corresponding non-periodic system \citep{Asboth2016short}.
In complete analogy, we introduce the time-quasimomentum $k_{t}$
for the Hamiltonian $\hat{H}_{t}$, so that the eigenstates of $\hat{H}_{t}$
are given by $\e^{\i k_{t}\alpha}\tau_{n,k_{t}}(\alpha)$ with $\tau_{n,k_{t}}(\alpha+2)=\tau_{n,k_{t}}(\alpha)$.
The first Chern numbers $\nu_{1}^{(t)}$ of the two bands in Fig.~\ref{fig:spectra}(b)
are then calculated by integrating the Berry curvature $\Omega_{k_{t}\varphi_{t}}$.
Note that by interpreting the phases $\varphi_{x}$ and $\varphi_{t}$
as quasimomenta, we increase the dimensionality of the systems. Each
of the Hamiltonians $\hat{H}_{x}$ and $\hat{H}_{t}$ thus describes
a 2D system, while their combination, $\hat{{\cal H}}'_{{\rm TB}}$,
whose spectrum is shown in Fig.~\ref{fig:spectra}(c), describes
a 4D system. The lowest and the highest bands are nondegenerate and
are characterized by the second Chern numbers calculated from the
Abelian Berry curvature \citep{PetridesEtAl2018,LeeEtAl2018}. Formally,
we gather the system parameters into a vector $\boldsymbol{R}=(k_{x},\varphi_{x},k_{t},\varphi_{t})$
and calculate the curvature as $\Omega_{\mu\nu}(\boldsymbol{R})=-2\Im\braket{\partial_{\mu}\xi_{n,k_{x},k_{t}}}{\partial_{\nu}\xi_{n,k_{x},k_{t}}}$
where $\partial_{\mu}\equiv\frac{\partial}{\partial R^{\mu}}$, $\mu=1,2,3,4$,
and $\ket{\xi_{n,k_{x},k_{t}}}$ is the cell-periodic part of the
$n$th band eigenstate of $\hat{{\cal H}}'_{{\rm TB}}$. Due to the
factorization $\ket{\xi_{n,k_{x},k_{t}}}=\ket{\chi_{n,k_{x}}}\otimes\ket{\tau_{n,k_{t}}}$,
the general formula for the second Chern number \citep{PetridesEtAl2018,LeeEtAl2018}
reduces to 
\begin{equation}
\nu_{2}^{(x,t)}=\frac{1}{4\pi^{2}}\int\d^{4}R\,\Omega_{k_{x}\varphi_{x}}\Omega_{k_{t}\varphi_{t}}=\nu_{1}^{(x)}\nu_{1}^{(t)}.\label{eq:nu2}
\end{equation}
The values of $\nu_{2}^{(x,t)}$ are indicated in Fig.~\ref{fig:spectra}(c).

Comparing the spectrum of $\hat{{\cal H}}'{}_{{\rm TB}}$ to the spectrum
of the exact tight-binding Hamiltonian $\hat{{\cal H}}{}_{{\rm TB}}$,
shown in Fig.~\ref{fig:spectra}(d), we note that they are nearly
identical. Slight discrepancies are to be expected since in order
to obtain the separable Hamiltonian we have neglected some very weak
couplings in $\hat{{\cal H}}{}_{{\rm TB}}$ \citep{supplement}. Nevertheless,
the second Chern numbers of the bands of energy spectra of $\hat{{\cal H}}{}_{{\rm TB}}$
and $\hat{{\cal H}}'{}_{{\rm TB}}$ are the same. This is supported
by the fact that the energy spectrum of $\hat{{\cal H}}_{{\rm TB}}$
can be obtained by adiabatically deforming the spectrum of $\hat{{\cal H}}'_{{\rm TB}}$
without closing the gaps in process. Relatedly, we remark that the
gap below the highest resonant energy band of $\hat{{\cal H}}_{{\rm TB}}$
remains open for all values of $\varphi_{x}$ and $\varphi_{t}$,
as shown in Fig.~\ref{fig:spectra}(e). The same is true for the
gap above the lowest band of $\hat{{\cal H}}_{{\rm TB}}$.

\emph{Higher-dimensional extensions. }Finally, let us consider an
optical lattice of two orthogonal spatial dimensions, so that the
full system Hamiltonian $\hat{H}_{4{\rm D}}=\hat{H}(x,\hat{p}_{x},t|\varphi_{x},\varphi_{t_{x}})+\hat{H}(y,\hat{p}_{y},t|\varphi_{y},\varphi_{t_{y}})$.
This produces a 4D time-space crystalline structure since the total
Wannier functions now have four independent indices: $W_{\boldsymbol{j},\boldsymbol{\alpha}}(x,y,t)=w_{j_{x},\alpha_{x}}(x,t)w_{j_{y},\alpha_{y}}(y,t)$,
where $\boldsymbol{j}=(j_{x},j_{y})$ and $\boldsymbol{\alpha}=(\alpha_{x},\alpha_{y})$
\citep{Zlabys2021}. A two-dimensional temporal structure of $2s\times2s$
sites now emerges in each two-dimensional spatial cell; motion in
the former is characterized by the temporal quasimomenta $k_{t_{x}}$
and $k_{t_{y}}$. The energy spectrum of this system may be readily
obtained as a Minkowski sum of two copies of spectra in Fig.~\ref{fig:spectra}(e).
The result is shown in Fig.~\ref{fig:spectra}(f), where it is apparent
that the highest and the lowest bands are separated from others by
a gap. This holds true not only for the displayed cut of the spectrum
at $\varphi_{x}=\varphi_{y}=\varphi_{t_{x}}=\varphi_{t_{y}}$, but
rather for all values of the phases. The ratio of the bandwidth of
the highest band to the gap below it is found to be 5\%, while the
ratio of the bandwidth of the lowest band to the gap above it is 2\%.

The system whose spectrum is shown in Fig.~\ref{fig:spectra}(f)
may thus be described by a lattice Hamiltonian
\begin{equation}
\hat{{\cal H}}_{8{\rm D}}=\hat{I}\otimes\hat{{\cal H}}_{{\rm TB}}^{(x)}+\hat{{\cal H}}_{{\rm TB}}^{(y)}\otimes\hat{I},
\end{equation}
where $\hat{I}$ is an identity matrix of the same size as $\hat{{\cal H}}_{{\rm TB}}$.
The system parameters are the two crystal momenta $k_{x}$, $k_{y}$,
the spatial phases $\varphi_{x}$ and $\varphi_{y}$, and the four
respective parameters of the two underlying temporal systems: $k_{t_{x}}$,
$k_{t_{y}}$, $\varphi_{t_{x}}$, $\varphi_{t_{y}}$. As in the 4D
case, the lowest and the highest energy bands are nondegenerate, and
therefore may be characterized by the fourth Chern number of a corresponding
Abelian gauge field. Generalizing (\ref{eq:nu2}) and related equations
to 8D in a straightforward way (see \citep{PetridesEtAl2018} and
\citep{supplement} for details), the relevant Chern number results
as $\nu_{4}^{(x,t_{x},y,t_{y})}=\nu_{2}^{(x,t_{x})}\nu_{2}^{(y,t_{y})}$.
This way we confirm that the highest and the lowest bands in Fig.~\ref{fig:spectra}(f)
are characterized by nonzero fourth Chern numbers, implying the topologically
nontrivial nature of the system. We note that if $\hat{{\cal H}}_{8{\rm D}}$
is constructed using two copies of the approximate Hamiltonian $\hat{{\cal H}}'_{{\rm TB}}$,
the higher gap closes, whereas the lower one remains open.

It is apparent in Fig.~\ref{fig:spectra}(f) that the highest and
the lowest bands are wider than the gaps, implying that the gaps disappear
if one more copy of the spectrum in Fig.~\ref{fig:spectra}(e) is
added. Nevertheless, a time-space structure based on a different spatial
system than the one given in (\ref{eq:h}) may exhibit even wider
gaps compared to those in Fig.~\ref{fig:spectra}(e). This would
allow one to realize a 12D time-space structure by combining three
copies of $\hat{{\cal H}}_{{\rm TB}}$, each based on a separate physical
dimension ($x$, $y$, and $z$).

\emph{Conclusions}. Summarizing, we have shown that the time-space
crystals may be used as a platform for studying 8D systems that can
be defined in a tight-binding form. We have devised a concrete, experimentally
realizable driven quantum system with validated parameters that is
an example of a topologically nontrivial 8D system. Remarkably, it
is possible to realize systems with nontrivial topological properties
and study the resulting effects in eight dimensions with the help
of a properly driven 2D system and without involving any internal
degrees of freedom of the particles. High-dimensional spatio-temporal
crystalline structures open up possibilities for building practical
devices that would be unthinkable in three dimensions. The results
presented in this Letter pave the way towards further research in
this direction.
\begin{acknowledgments}
This research was funded by the National Science Centre, Poland, Project
No.~2021/42/A/ST2/00017 (K.\,S.) and the Lithuanian Research Council,
Lithuania, Project No.~S-LL-21-3. For the purpose of Open Access,
the authors have applied a CC-BY public copyright license to any Author
Accepted Manuscript (AAM) version arising from this submission.
\end{acknowledgments}

\end{document}

\global\long\def\i{\mathrm{i}}%
\global\long\def\e{\mathrm{e}}%
\global\long\def\d{\mathrm{d}}%
\global\long\def\bra#1{\left\langle #1\right|}%
\global\long\def\ket#1{\left|#1\right\rangle }%
\global\long\def\braket#1#2{\left\langle #1|#2\right\rangle }%
\global\long\def\ketbra#1#2{\left|#1\right\rangle \!\left\langle #2\right|}%
\global\long\def\Tr{\mathrm{Tr}}%

\title{Supplementary Material for:\\
\textcolor{black}{Eight-dimensional topological systems simulated
using }time-space crystalline structures}
\author{Yakov Braver}
\affiliation{Institute of Theoretical Physics and Astronomy, Vilnius University,
Saul\.{e}tekio 3, LT-10257 Vilnius, Lithuania}
\author{Egidijus Anisimovas}
\affiliation{Institute of Theoretical Physics and Astronomy, Vilnius University,
Saul\.{e}tekio 3, LT-10257 Vilnius, Lithuania}
\author{Krzysztof Sacha}
\affiliation{Instytut Fizyki Teoretycznej, Uniwersytet Jagiello\'{n}ski, ulica
Profesora Stanis\l awa \L ojasiewicza 11, PL-30-348 Krak\textbf{\'{o}}w,
Poland}
\maketitle

\section{Diagonalization of Floquet Hamiltonian}

In this section, we sketch the diagonalization of the Floquet Hamiltonian
\begin{equation}
\hat{{\cal H}}_{k}=\hat{H}_{k}-\i\partial_{t}.
\end{equation}
Here, we have introduced the quasimomentum index, stemming from the
spatial part of the problem. Specifically, we have

\begin{equation}
\hat{H}_{k}(x,t|\varphi_{t})=\hat{h}_{k}(x)+\xi_{{\rm S}}(x,t)+\xi_{{\rm L}}(x,t|\varphi_{t})\label{eq:H}
\end{equation}
where

\begin{equation}
\hat{h}_{k}=(\hat{p}_{x}+k)^{2}+V\sum_{n=0}^{3N}\delta(x-\tfrac{na}{3})+U\sum_{n=0}^{2}g_{n}(x)\cos(\varphi_{x}+\tfrac{2\pi n}{3})\label{eq:h}
\end{equation}
with 
\begin{equation}
g_{n}(x)=\begin{cases}
1, & \frac{n}{3}a\leq x\bmod a<\frac{n+1}{3}a,\\
0, & \text{otherwise},
\end{cases}
\end{equation}
and

\vspace{-5mm}
\begin{subequations}
\label{eq:perturb}
\begin{align}
\xi_{{\rm S}}(x,t) & =\lambda_{{\rm S}}\cos(\tfrac{12\pi x}{a})\cos(2\omega t),\\
\xi_{{\rm L}}(x,t|\varphi_{t}) & =\lambda_{{\rm L}}\cos(\tfrac{6\pi x}{a})\cos(\omega t+\varphi_{t}).
\end{align}
\end{subequations}

It is a standard exercise to obtain the Bloch modes $\psi_{m,k}$
and eigenenergies $E_{m,k}$ for a delta-function Hamiltonian (\ref{eq:h}).
The next step is to solve the eigenvalue problem
\begin{equation}
\hat{{\cal H}}_{k}v_{n,k}(x,t)=\varepsilon_{n,k}v_{n,k}(x,t).
\end{equation}
Instead of the full diagonalization of the Floquet Hamiltonian $\hat{{\cal H}}_{k}$,
our aim is to obtain an effective (secular) time-independent Hamiltonian
which can be more easily diagonalized. To this end we perform a time-dependent
unitary transformation 
\begin{equation}
\psi'_{m,k}(x,t)=\e^{-\i\nu(m)\omega t/s}\psi_{m,k}(x).\label{eq:psi'}
\end{equation}
Here, the function $\nu(m)=\lceil m/3\rceil$ (where $\lceil\cdots\rceil$
is the ceiling operation) transforms the level numbers $m=1,2,3,4,5,6,\ldots$
into band indices
\begin{equation}
\nu=1,1,1,2,2,2,3,3,3,\ldots\label{eq:nu}
\end{equation}
Note that this labeling is correct as long as we keep $U$ small enough
so that the difference of depths of the sites is always smaller than
the gaps between the energy bands of the unperturbed Hamiltonian.

The diagonal matrix elements of $\hat{{\cal H}}_{k}$ are 
\begin{equation}
\bra{\psi'_{m',k}}\left(\hat{h}_{k}-\i\frac{\partial}{\partial t}\right)\ket{\psi'_{m,k}}=\left(E_{m,k}-\frac{\nu(m)\omega}{s}\right)\delta_{m',m}.\label{eq:diag}
\end{equation}
For the perturbation term $\xi_{{\rm L}}$, we obtain
\begin{equation}
\begin{split}\bra{\psi'_{m',k}}\xi_{{\rm L}}\ket{\psi'_{m,k}} & =\lambda_{{\rm L}}\cos(\omega t+\varphi_{t})\,\e^{-\i(\omega/s)[\nu(m)-\nu(m')]t}\\
 & \quad\times\intop_{0}^{a}\d x\,\psi_{m',k}^{*}(x)\cos\tfrac{6\pi x}{a}\psi_{m,k}(x).
\end{split}
\end{equation}
 Applying the secular approximation, we replace
\begin{equation}
\cos(\omega t+\varphi_{t})\,\e^{-\i(\omega/s)[\nu(m)-\nu(m')]t}
\end{equation}
with its time-independent contribution

\begin{equation}
\tfrac{1}{2}(\e^{\i\varphi_{t}}\delta_{\nu'+s,\nu}+\e^{-\i\varphi_{t}}\delta_{\nu'-s,\nu}),
\end{equation}
where $\nu'\equiv\nu(m')$. Similarly, we have
\begin{equation}
\begin{split}\bra{\psi'_{m',k}}\xi_{{\rm S}}\ket{\psi'_{m,k}} & =\lambda_{{\rm S}}\cos(2\omega t)\,\e^{-\i(\omega/s)[\nu(m)-\nu(m')]t}\\
 & \quad\times\intop_{0}^{a}\d x\,\psi_{m',k}^{*}(x)\cos\tfrac{12\pi x}{a}\psi_{m,k}(x).
\end{split}
\end{equation}
where we replace
\begin{equation}
\cos(2\omega t)\,\e^{-\i(\omega/s)[\nu(m)-\nu(m')]t}
\end{equation}
with

\begin{equation}
\tfrac{1}{2}(\delta_{\nu'+2s,\nu}+\delta_{\nu'-2s,\nu}),
\end{equation}

\begin{figure}
\begin{centering}
\includegraphics{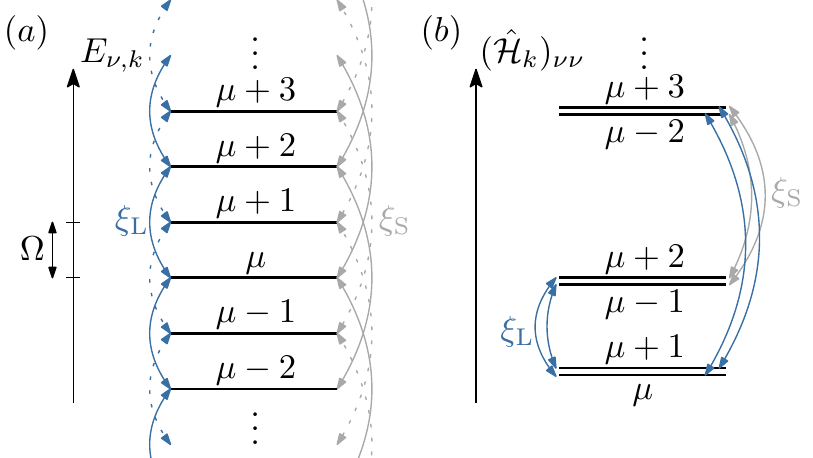}
\par\end{centering}
\caption{\label{fig:coupling}Schematic representation of the perturbation-induced
band coupling (distances between bands not drawn to scale). (a) Eigenenergy
bands of the unperturbed Hamiltonian $\hat{h}_{k}$ in the vicinity
of a certain resonant band whose number is $\mu$. Each horizontal
line represents an energy band (which contains three subbands) (b)
Diagonal elements of Floquet Hamiltonian represented as coupled quasienergy
levels. Only the couplings that have the most effect are shown.}
\end{figure}

Figure \ref{fig:coupling}(a) displays schematically the coupling
of the eigenenergy bands of $\hat{h}_{k}$ induced by the perturbations.
Each band of $\hat{h}_{k}$ consists of three subbands because each
cell of the spatial potential of $\hat{h}_{k}$ contains three sites,
but this is not detailed in the figure. As a result of the secular
approximation, each level of the $\nu$th band is coupled to all the
levels of bands $\nu\pm s$ (via $\xi_{{\rm L}}$) and $\nu\pm2s$
(via $\xi_{{\rm S}}$); in our case, $s=2$. The driving frequency
is chosen as $\omega=\Omega s$ with $\Omega$ being the energy gap
between certain bands $\mu$ and $\mu+1$. In practice, however, we
fine-tune $\omega_{\mu}$ to get the optimal quasienergy spectrum.
Below, we will assume the working point of $\mu=30$, which yields
$E_{31}-E_{30}\approx340$ for $a=4$. This is consistent with the
result calculated for the problem of a particle in an infinitely deep
potential well of width $l=a/3$: in recoil units, the energy spectrum
is $E_{\nu}=(\nu\pi/l)^{2}$. This also allows one to estimate the
frequency: $\Omega=E_{\mu+1}-E_{\mu}=(2\mu+1)\frac{\pi^{2}}{l^{2}}$.
In a small neighborhood of the $\mu$th energy band, the spectrum
is close to being equidistant, therefore, transitions such as $E_{\mu-1}\leftrightarrow E_{\mu+1}$
and $E_{\mu+1}\leftrightarrow E_{\mu+3}$ are almost resonant with
the frequency $\omega$.

A sketch of the quasienergy spectrum of $\hat{{\cal H}}_{k}$ is given
in Fig.~\ref{fig:coupling}(b). The bands of $\hat{h}_{k}$ become
grouped into overlapping pairs, and the resonant bands are the lowest.
This can be seen by employing the above estimates of the energy spectrum
$E_{\nu}$ and the resonant frequency $\Omega$. Suppressing the $k$
index, we find from Eq.~(\ref{eq:diag}) that
\begin{equation}
{\cal H}_{\mu+\beta,\mu+\beta}=E_{\mu+\beta}-\Omega(\mu+\beta)={\cal H}_{\mu,\mu}+(\beta^{2}-\beta)\tfrac{\pi^{2}}{l^{2}},\label{eq:H-beta}
\end{equation}
where integer (band) index $\beta$ runs from $1-\mu$ to infinity.
Assuming the perturbation is small, the above expression yields the
Floquet quasienergy spectrum shown in Fig.~\ref{fig:coupling}(b).
The lowest are the levels $\mu$ and $\mu+1$, whose energies should
coincide according to Eq.~(\ref{eq:H-beta}).

The actual quasienergy spectrum is shown in Fig.~\ref{fig:2D-pump}(a)
for the case of simultaneous time-space pumping ($\varphi_{x}=\varphi_{t}$),
with only the resonant bands displayed. These correspond precisely
to the bands $\mu-1$ through $\mu+2$ shown schematically in Fig.~\ref{fig:coupling}(b).

Returning to the problem of finding the eigenfunctions of $\hat{{\cal H}}_{k}$,
it remains to numerically diagonalize the obtained matrix for each
$k$ to find the coefficients $b_{m,k}^{(n)}$ of the expansion

\begin{equation}
v_{n,k}(x,t)=\sum_{m=1}^{\infty}b_{m,k}^{(n)}\psi'_{m,k}(x,t).
\end{equation}

All calculations have been performed using a number of software packages
\citep{JuliaDiffEq,JuliaDiffEq2,JuliaDiffEqKahanLi,JuliaDiffEqMcAte,JuliaOptim,JuliaIntervalArithmetic}
written in the Julia programming language \citep{Julia}. The source
code of our package is available on GitHub \citep{package}.

\section{Visualization of time-space pumping using the Wannier functions\label{sec:Visualization-of-time-space}}

\begin{figure*}
\begin{centering}
\includegraphics[width=17.2cm]{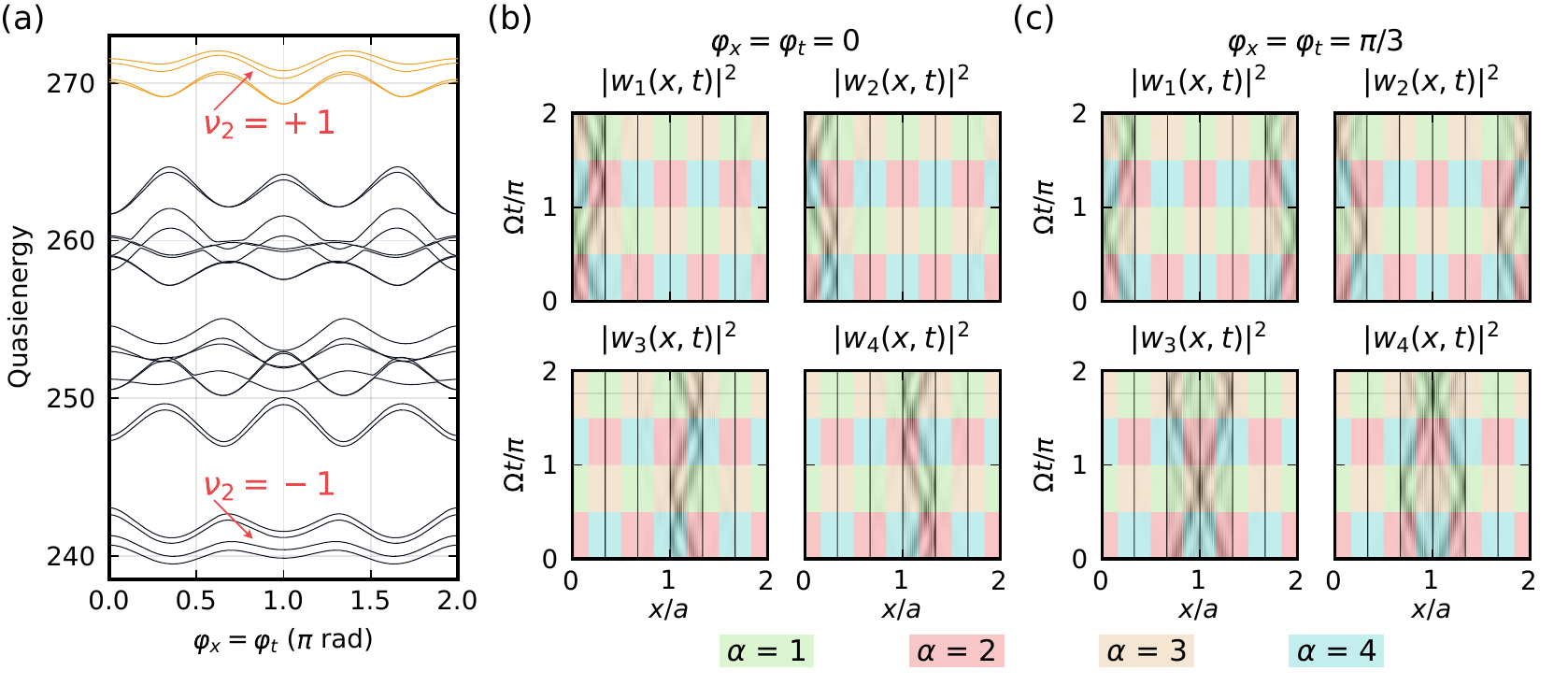}
\par\end{centering}
\caption{\label{fig:2D-pump}Time-space adiabatic pumping in a 2D time-space
crystal with $s=2$ temporal cells and $N=2$ spatial cells that consists
of three spatial sites. The following values of parameters were used:
$a=4.000$, $V=2000$, $U=7.000$, $\omega=676.8$, $\lambda_{{\rm S}}=10.00$,
$\lambda_{{\rm L}}=20.00$. (a) Quasienergy spectrum of the Floquet
Hamiltonian $\hat{{\cal H}}$. (b) The Wannier functions $|w_{\ell}(x,t)|^{2}$
at $\varphi_{x}=\varphi_{t}=0$, represented by black regions. The
shaded areas indicate the extent of the temporal sites numbered by
$\alpha$: sites $\alpha=1,2$ belong to the first temporal cell,
while $\alpha=3,4$ belong to the second. Black vertical lines separate
the spatial sites. (c) Same as (b) at $\varphi_{x}=\varphi_{t}=\pi/3$.}
\end{figure*}
To analyze the pumping dynamics, we construct Wannier functions $w_{\ell}(x,t)$
by diagonalizing the periodic position operator $\e^{\i\frac{2\pi}{Na}x}$
\citep{Asboth2016short,Aligia1999}. The calculation is performed
in the basis of functions $u_{n,k}(x,t)=\e^{\i kx}v_{n,k}(x,t)$,
where we have complemented the Floquet--Bloch modes $v_{n,k}$ by
the exponential factor to obtain the ``full'' wave function defined
on $x\in[0;Na)$. Diagonalization yields the coefficients $d_{n,k}^{(\ell)}$
of the expansion
\begin{equation}
w_{\ell}(x,t)=\sum_{n,k}d_{n,k}^{(\ell)}u_{n,k}(x,t).\label{eq:wl}
\end{equation}
Diagonalization has to be performed at a single fixed time moment
at which the modes $u_{n,k}$ do not overlap strongly so that the
resulting Wannier functions exhibit the required temporal localization
\citep{Zlabys2021}. In the relevant cases we sum over all quasimomenta
$k$, while index $n$ runs over a selected range of modes. Figure
\ref{fig:2D-pump}(b) displays the functions $w_{\ell}$ at $\varphi_{x}=\varphi_{t}=0$
constructed by mixing the four modes corresponding to the highest
band in the spectrum in Fig.~\ref{fig:2D-pump}(a) (highlighted in
orange). The resulting functions inherit the spatial localization
of the Floquet--Bloch modes being mixed and, consequently, are confined
to a single spatial site. A detector placed at one of the turning
points of any spatial site will be registering periodic arrival of
a particle, and it is precisely this periodicity that allows us to
speak of a crystalline structure in time \citep{SachaTC2020,Zlabys2021,Braver2022}.
The shaded regions in Fig.~\ref{fig:2D-pump}(b) indicate the parts
of the time-space that we associate --- by convention --- with the
four temporal sites: sites $\alpha=1,2$ (green and red) belong to
the first temporal cell, while $\alpha=3,4$ (beige and blue) belong
to the second. For example, a particle in the state $w_{1}$ will
be passing a detector placed at $x=0$ at time intervals $(\Omega t\bmod2\pi)\in[0,\pi]$,
meaning that it occupies the first temporal cell. In the course of
the pumping, the states transition both in the spatial and in the
temporal directions. At $\varphi_{x}=\varphi_{t}=\pi/3$ {[}see Fig.~\ref{fig:2D-pump}(c){]},
the states are localized in a single temporal site while occupying
two neighboring spatial sites. At the end of the pumping cycle, states
$w_{1}$ and $w_{4}$ effectively exchange their initial occupations,
as do states $w_{2}$ and $w_{3}$ (not shown). Thus, having divided
the time-space into temporal and spatial cells as shown in the figure,
we find that at the end of a time-space pumping cycle each Wannier
function occupies spatial and temporal cells different from the starting
ones.

\section{The Wannier basis and the tight-binding picture}

\begin{figure*}
\begin{centering}
\includegraphics[width=17.2cm]{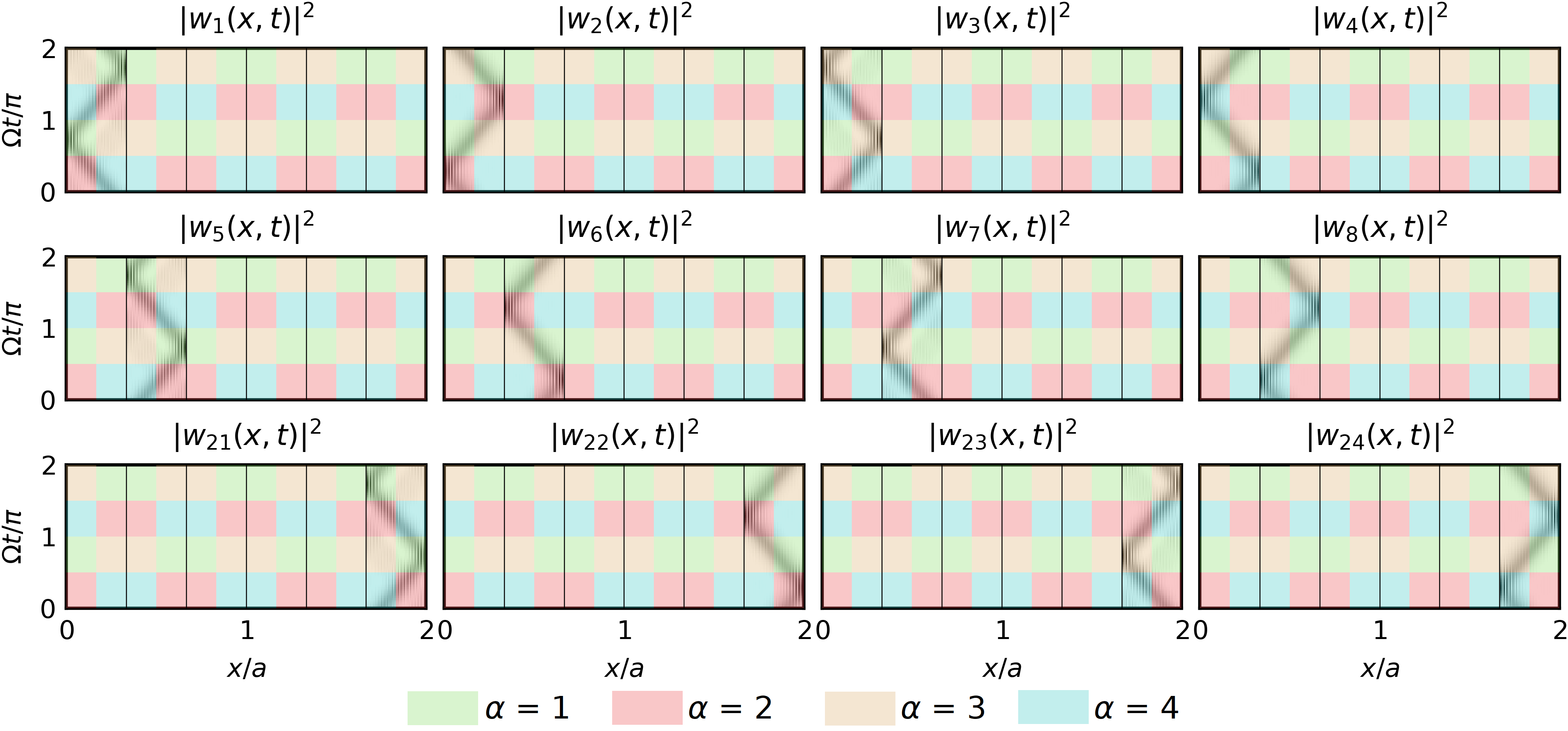}
\par\end{centering}
\caption{\label{fig:wanniers}Twelve Wannier functions.}
\end{figure*}
 To switch to the tight-binding description, we construct the Wannier
basis functions by mixing the states corresponding to all the resonant
levels of the Floquet quasienergy spectrum, i.e., all levels shown
in Fig.~\ref{fig:2D-pump}(a). In practice, two out of four Wannier
functions localized in a single spatial site come out not well-localized
in the temporal dimensions. To alleviate this issue, we perform an
additional unitary transformation of the two problematic Wannier functions
$w_{a}$ and $w_{b}$:
\begin{equation}
\binom{\tilde{w}_{a}}{\tilde{w}_{b}}=\hat{U}(\gamma,\theta,\phi)\binom{w_{a}}{w_{b}}.\label{eq:optimise}
\end{equation}
Here, $\hat{U}(\gamma,\theta,\phi)$ is a general $2\times2$ unitary
matrix
\begin{equation}
\hat{U}(\gamma,\theta,\phi)=\e^{\i\gamma\boldsymbol{n}(\theta,\phi)\boldsymbol{\sigma}}
\end{equation}
where $\gamma\in[0,2\pi]$, $\boldsymbol{\sigma}$ is a vector of
Pauli matrices, and $\boldsymbol{n}$ is a three-component unit vector
parameterized by an azimuthal angle $\phi\in[0,2\pi]$ and a polar
angle $\theta\in[0,\pi]$. We optimize the angles $\gamma$, $\theta$,
and $\phi$ to get the least possible overlap of probability densities
of $\tilde{w}_{a}$ and $\tilde{w}_{b}$. Specifically,
\begin{equation}
\iint\d x\,\d t\,\left||\tilde{w}_{3}|^{2}-|\tilde{w}_{2}|^{2}\right|\to{\rm max}.
\end{equation}
In what follows, the Wannier functions being discussed are the ``optimized''
ones.

\begin{figure}
\begin{centering}
\includegraphics{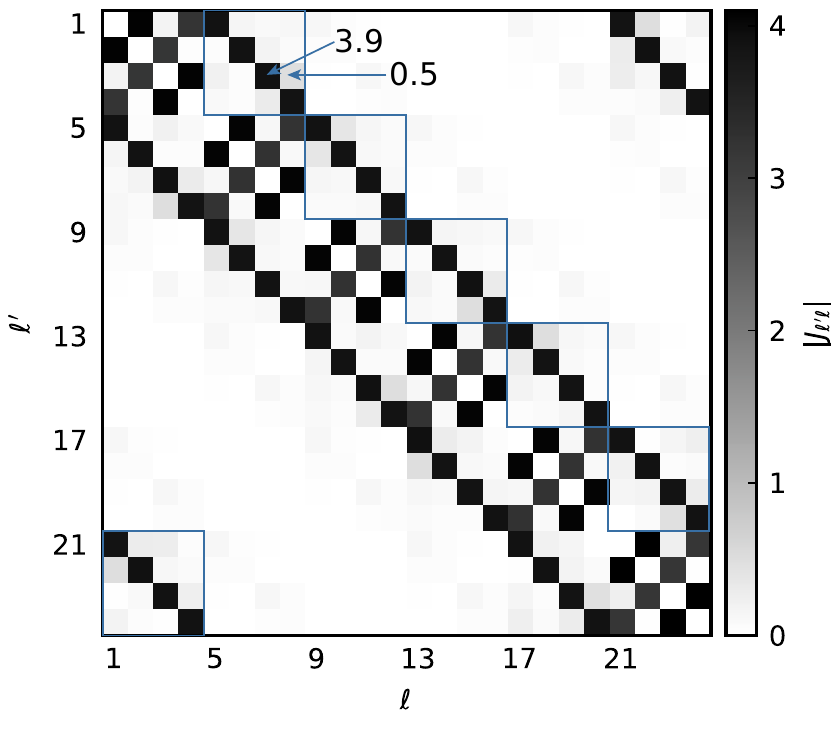}
\par\end{centering}
\caption{\label{fig:H-TB}Absolute values $|J_{\ell'\ell}|$ of the matrix
elements of $\hat{{\cal H}}_{{\rm TB}}$ at $\varphi_{x}=\varphi_{t}=0$.
Diagonal elements have been set to zero for visual clarity. Values
of $J_{37}$ and $J_{38}$ are explicitly indicated. The structure
of the blue blocks is similar to the displayed one for all values
of the phases $\varphi_{x}$ and $\varphi_{t}$.}
\end{figure}

In the studied case, we obtain $3N\times2s=24$ Wannier functions,
which we are free to number as we see fit. Figure \ref{fig:wanniers}
shows, as an example, twelve Wannier functions localized in the first,
second, and sixth spatial sites As we can see, each of them is strongly
localized in a single spatial and temporal site. The numbering convention
is such that the $\ell$th Wannier is localized in the $j$th spatial
and the $\alpha$th temporal sites, with $\ell=2s(j-1)+\alpha$, where
$j\in[1,3N]$ and $\alpha\in[1,2s]$. Note, however, that the temporal
sites are assigned differently in the odd and even spatial sites.
For example, a particle is said to occupy the first temporal site
(green areas in Fig.~\ref{fig:wanniers}) of an \emph{odd} spatial
site if it will most likely be found at the \emph{right} turning point
in the interval $(\Omega t\bmod2\pi)\in[3\pi/2,2\pi]$, see $w_{1}$
in Fig.~\ref{fig:wanniers}. On the other hand, a particle occupying
the first temporal site of an \emph{even }spatial site will most likely
be found found at the \emph{left }turning point in the same interval
$(\Omega t\bmod2\pi)\in[3\pi/2,2\pi]$, see $w_{5}$ in Fig.~\ref{fig:wanniers}.
Such a convention ensures the correct interpretation of the outcome
of a pumping cycle, as demonstrated in Section \ref{sec:Visualization-of-time-space}.
Moreover, it allows one to obtain a tight-binding Hamiltonian 
\begin{equation}
\hat{{\cal H}}_{{\rm TB}}(\varphi_{x},\varphi_{t})=\sum_{\ell',\ell}J_{\ell'\ell}(\varphi_{x},\varphi_{t})\hat{a}_{\ell'}^{\dagger}\hat{a}_{\ell}
\end{equation}
that can be decomposed into two independent parts --- the spatial
Hamiltonian and the temporal one. The absolute values of the matrix
elements of $\hat{{\cal H}}_{{\rm TB}}$ calculated at $\varphi_{x}=\varphi_{t}=0$
are presented in Fig.~\ref{fig:H-TB}, where the diagonal elements
have been set to zero for visual clarity. It is apparent that the
resulting matrix allows for the decomposition
\begin{equation}
\hat{{\cal H}}_{{\rm TB}}\approx\hat{I}_{x}\otimes\hat{H}_{t}+\hat{H}_{x}\otimes\hat{I}_{t}\equiv\hat{{\cal H}}_{{\rm TB}}^{\prime},\label{eq:separate}
\end{equation}
as discussed in the Letter. In our case, we take $\hat{H}_{t}$ to
be given by the top left $4\times4$ block of $\hat{{\cal H}}_{{\rm TB}}$,
while $\hat{H}_{x}$ is given by
\begin{equation}
\hat{H}_{x}=\begin{pmatrix}J_{11} & |J_{15}| & 0 & 0 & 0 & |J_{15}|\\
|J_{15}| & J_{55} & |J_{15}| & 0 & 0 & 0\\
0 & |J_{15}| & J_{99} & |J_{15}| & 0 & 0\\
0 & 0 & |J_{15}| & J_{11} & |J_{15}| & 0\\
0 & 0 & 0 & |J_{15}| & J_{55} & |J_{15}|\\
|J_{15}| & 0 & 0 & 0 & |J_{15}| & J_{99}
\end{pmatrix}.
\end{equation}
For simplicity, we assume that the phases of the the tunneling strengths
have been eliminated by redefining the global phases of the basis
functions. The crudest part of the separability approximation (\ref{eq:separate})
is the dropping of the off-diagonal elements in the blue blocks (and
their conjugate counterparts) highlighted in Fig.~\ref{fig:H-TB}.

\begin{figure}
\begin{centering}
\includegraphics{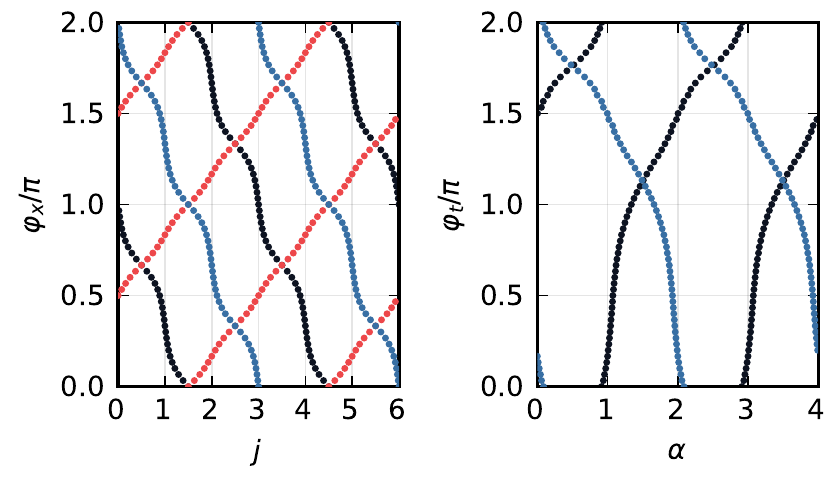}
\par\end{centering}
\caption{\label{fig:pump}Dynamics of the Wannier centers. (a) Spatial pumping
dynamics calculated based on $\hat{H}_{x}$. Black, red, and blue
points represent, respectively, the positions of the Wannier centers
calculated by mixing the eigenstates corresponding to the bottom,
middle, and top bands in the spectrum of $\hat{H}_{x}$ {[}see Fig.~2(a)
in the Letter{]}. (b) Temporal pumping dynamics calculated based on
$\hat{H}_{t}$. Black and blue points represent, respectively, the
positions of the Wannier centers calculated by mixing the eigenstates
corresponding to the bottom and top bands in the spectrum of $\hat{H}_{t}$
{[}see Fig.~2(b) in the Letter{]}.}
\end{figure}
To extract the first Chern numbers $\nu_{1}$ of the bands of Hamiltonians
$\hat{H}_{x}$ and $\hat{H}_{t}$, we may calculate the pumping dynamics.
To this end, we consider the periodic coordinate operator, whose elements,
in coordinate representation, are given by $X_{mm}=\e^{\i\frac{2\pi}{M}m}$,
where $m\in[0,M-1]$ and $M$ is the size of the Hamiltonian under
consideration ($M=6$ for $\hat{H}_{x}$ and $M=4$ for $\hat{H}_{t}$).
Then, we diagonalize this operator in the subspace of the eigenstates
of $\hat{H}_{x}$ or $\hat{H}_{t}$ corresponding to a single band
of the energy spectrum. The resulting eigenfunctions of $\hat{X}$
are the Wannier functions, while the phases of the eigenvalues (scaled
by $2\pi/M$) are their well-defined positions (also called the Wannier
centers) \citep{Aligia1999,Asboth2016short}. Black points in Fig.~\ref{fig:pump}(a)
show the dynamics of the Wannier centers obtained by considering the
subspace of the lowest energy band of $\hat{H}_{x}$. It is apparent
that the states are transferred by one cell (three sites) ``to the
left'' during one pumping cycle. Blue points show an analogous result
for the states of the highest energy band of $\hat{H}_{x}$, while
the red points indicate a displacement by two cells per cycle in the
opposite direction. The first Chern numbers of the bands are thus
$-1$, $+2$, and $-1$ for the bottom, middle, and the top energy
bands of $\hat{H}_{x}$.

The temporal pumping based on $\hat{H}_{t}$ is studied in Fig.~\ref{fig:pump}(b).
The black points indicate a displacement by a single cell ``to the
right'' per pumping cycle taking place in the bottom energy band
($\nu_{1}=+1$), while the blue points show a displacement by a single
cell in the opposite direction for the top energy band ($\nu_{1}=-1$).

\section{The fourth Chern number}

As discussed in the Letter, combining two 4D systems described by
$\hat{{\cal H}}_{{\rm TB}}$ and constructed in orthogonal spatial
dimensions yields an 8D system according to 
\begin{equation}
\hat{{\cal H}}_{8{\rm D}}=\hat{I}\otimes\hat{{\cal H}}_{{\rm TB}}^{(x)}+\hat{{\cal H}}_{{\rm TB}}^{(y)}\otimes\hat{I}.\label{eq:H_8D}
\end{equation}
The parameters of $\hat{{\cal H}}_{8{\rm D}}$ are the components
of the following formal vector: 
\begin{equation}
\boldsymbol{R}=(k_{x},\varphi_{x},k_{t_{x}},\varphi_{t_{x}},k_{y},\varphi_{y},k_{t_{y}},\varphi_{t_{y}}).
\end{equation}
The Abelian Berry curvature is introduced using the eigenstates of
a nondegenerate energy band of $\hat{{\cal H}}_{8{\rm D}}$ as 
\begin{equation}
\Omega_{\mu\nu}(\boldsymbol{R})=-2\Im\braket{\partial_{\mu}\zeta_{n,\boldsymbol{k}}}{\partial_{\nu}\zeta_{n,\boldsymbol{k}}}
\end{equation}
where $\partial_{\mu}\equiv\frac{\partial}{\partial R^{\mu}}$, index
$\mu$ runs from 1 through 8, $\boldsymbol{k}=(k_{x},k_{y},k_{t_{x}},k_{t_{y}}$),
and $\ket{\zeta_{n,\boldsymbol{k}}}$ is the cell-periodic part of
the $n$th band eigenstate of $\hat{{\cal H}}_{8{\rm D}}$. Let $\ket{n,k_{x},k_{t_{x}}}$
be the cell-periodic part of the $n$th band eigenstate of $\hat{{\cal H}}_{{\rm TB}}^{(x)}$;
then (\ref{eq:H_8D}) implies $\ket{\zeta_{n,\boldsymbol{k}}}=\ket{n,k_{x},k_{t_{x}}}\otimes\ket{n,k_{y},k_{t_{y}}}$.

The fourth Chern number is given by \citep{PetridesEtAl2018}
\begin{equation}
\nu_{4}=\frac{1}{(2\pi)^{4}}\int\d^{8}R\,\frac{\epsilon_{\mu_{1},\ldots,\mu_{8}}}{(2^{4}\times4!)}\Omega_{\mu_{1}\mu_{2}}\Omega_{\mu_{3}\mu_{4}}\Omega_{\mu_{5}\mu_{6}}\Omega_{\mu_{7}\mu_{8}},
\end{equation}
where $\epsilon_{\mu_{1},\ldots,\mu_{8}}$ is the totally antisymmetric
tensor of rank 8. Due to the separability of the system, this expression
simplifies to
\begin{equation}
\nu_{4}=\frac{1}{(2\pi)^{4}}\int\d^{8}R\,\Omega_{12}\Omega_{34}\Omega_{56}\Omega_{78}=\nu_{2}^{(x,t_{x})}\nu_{2}^{(y,t_{y})}.
\end{equation}

\bibliographystyle{aipnum4-2}
%aipnum4-2.bst 2019-01-14 (MD) hand-edited version of apsrev4-1.bst